\documentclass[12pt]{article}

\usepackage{mathtools, cuted}
\usepackage{paralist}
\usepackage{subcaption}
\usepackage{stfloats}
\usepackage{hyperref}

\usepackage{setspace}
\usepackage{apacite}
\usepackage{natbib}

\title{Modeling of Wastewater Treatment Processes with HydroSludge}

\author{Sergio Iserte$^1$ \and Pablo Carratalà$^1$ \and Rosario Arnau$^1$ \and Raúl Martínez-Cuenca$^1$ \and Paloma Barreda$^1$ \and Luís Basiero$^2$ \and Javier Climent$^{1,2}$ \and Sergio Chiva$^1$\thanks{Corresponding author: schiva@uji.es}}
\date{%
    $^1$Universitat Jaume I (UJI), Spain\\%
    $^2$Sociedad Fomento Agrícola Castellonense (FACSA), Spain\\[2ex]%
}

\usepackage{xcolor,colortbl}
\newcommand{\review}[1]{\textcolor{black}{#1}}
\newcommand{\reviewMinor}[1]{\textcolor{black}{#1}}
\usepackage[multiple]{footmisc}

\begin{document}

\maketitle
%\input{sections/sections-template}

%https://www.elsevier.com/journals/environmental-modelling-and-software/1364-8152/guide-for-authors

\textbf{Abstract:}
The pressure for Water Resource Recovery Facilities (WRRF) operators to efficiently treat wastewater is greater than ever because of the water crisis, produced by the climate change effects and more restrictive regulations. Technicians and researchers need to evaluate WRRF performance to ensure maximum efficiency. For this purpose, numerical techniques, such as CFD, have been widely applied to the wastewater sector to model biological reactors and secondary settling tanks with high spatial and temporal accuracy. \review{However, limitations such as complexity and learning curve, prevent extending CFD usage among wastewater modeling experts.} This paper presents HydroSludge, a framework that provides a series of tools that simplify the implementation of the processes and workflows in a WRRF. This work leverages HydroSludge to preprocess existing data, aid the meshing process, and perform CFD simulations. Its intuitive interface proves itself as an effective tool to increase the efficiency of wastewater treatment.
\newline

\textbf{Keywords:} OpenFOAM; Submersible Mixers; Meshing Assistance; ASM1; Parallel Computing.

\section{Introduction}
In recent years, initiatives related to the protection of natural resources such as water have undergone significant growth worldwide. When it comes to water, the quality of the effluent to be discharged by Water Resource Recovery Facilities (WRRF) is becoming more and more restrictive. It is necessary to maintain the pace of improvement and efficiency of the different processes in WRRF. Notably, biological treatment is one of the most critical tasks in a WRRF~\citep{Grady2011}. \review{The biological treatment involves biological reactors and the Secondary Settling Tanks (SST). For this reason, it can represent around 50\% of the whole plant's total energy consumption, depending on the configuration and the aeration system used~\citep{Ovezea2009}.}

Mathematical modeling and numerical simulation have become essential for most of the physical and biochemical processes in WRRFs. These techniques are adopted as great potential tools during the design, operation, optimization, and process control of a WRRF ~\citep{JacekMakinia2010}. The main reason for using models and simulations is to enhance efficiency and save costs on resources exploring the dynamic behavior of the process units under different conditions. 
Software platforms such as Biowin, GPS-X, WEST, Simba, or SUMO implement different numerical methods and are well accepted in the water sector because they are relatively simple to manage and provide global results quickly. 
However, their hydrodynamics modeling approach does not account for hydraulic phenomena in real performance: geometry of the tanks, internal elements, gradient of the state variables, and the spatial distribution inside the tanks. Furthermore, they cannot detect mixing problems such as short-circuiting, dead volumes, or stratification/inhomogeneity, which commonly occur in real tanks, and play a vital role in the pollutant removal efficiency~\citep{Samstag2016}.

Computational Fluid Dynamics (CFD) represents the most sophisticated mechanism of simulation able to reproduce in detail the fluid behavior and to analyze specific hydraulic troubleshooting in 3D. Concretely, many of the most essential characteristics in wastewater treatment modeling can be included in a CFD. For instance, multi-phase flow, biokinetics, and submodels such as aeration, sedimentation, or non-Newtonian fluids. Despite the notable increase in scientific production related to CFD modeling since its adoption into wastewater treatment~\citep{Glover2006}, the main bottleneck to widespread the use of the CFD in this field is the learning curve and the cost of the commercial software license. To address these issues, we have designed, developed, and evaluated a software named HydroSludge. \reviewMinor {Notice that users should still be aware of which fluid model they need for their simulation.}

HydroSludge stems from the necessity of providing a specific simulation software for WRRF management. For this purpose, HydroSludge is presented as an easy--to--use framework for the water sector. With HydroSludge, authors aim to extend the use of CFD technologies among water treatment professionals by reducing the operation's complexity. HydroSludge provides holistic tools for the CFD workflow (data pre-processing, geometry, meshing, solver, and post-processing). Besides, it is designed to leverage parallelism and provide higher performance by leveraging all the underlying resources available on the host computer. 

The rest of the paper is structured as follows:
Section~\ref{sec:methods} introduces and describes the methods implemented in HydroSludge.
Section~\ref{sec:software} provides software details and outlines the modules that compose HydroSludge. 
Section~\ref{sec:results}, presents a series of use cases to test and evaluate the platform. Finally, Section~\ref{sec:conclusions} contains the conclusions of this research and development.
%, as well as, defines future interesting future work.

\section{Methods}\label{sec:methods}
In this section, the most relevant methods provided by HydroSludge are introduced and detailed.
The section is split into two parts. On the one hand, the tools for filtering and checking the input data are described in Section~\ref{subsec:preliminary}.
On the other hand, the means for building a mesh and performing simulations are detailed in Section~\ref{subsec:simulation}.

\subsection{Preliminary Study}\label{subsec:preliminary}
To provide appropriate input to a CFD simulation, the collected data must follow quality standards. In this regard, HydroSludge facilitates the filtering and checking operations for the input data.
\subsubsection{Filtering}\label{subsubsec:filter}
This operation is composed of three independent stages:
\begin{itemize}
\item{Settling}
The drift flux model~\citep{Brennan2001} is one of the most popular among SST-CFD studies in the literature. It is commonly used to calculate the two-phase mixture of the mixed liquor. For this reason, HydroSludge provides a tool to calculate the activated sludge solids settling velocity through the experimental batch settlings data.
In order to reproduce the settling velocity for clarifiers and fit the constant terms automatically, HydroSludge includes the model of~\cite{vesilind1968design}
\begin{equation}
V_s = V_0 e^{-r_hX}
\label{eq:vesilin}
\end{equation}
and the model of~\cite{Takacs}
\begin{equation}
V_s=V_0(e^{-r_hX}-e^{-r_pX})
\label{eq:takacs}
\end{equation}
(eq. \ref{eq:takacs}) models. Where $V_s$ is the settling velocity, $V_0$ is the settling velocity when the total solids concentration tends to zero, and $r_h$ and $r_p$, are the characteristic empirical parameters of the sludge calculated from the batch experiments.
\item{Inflow}
WRRFs are dynamic systems that can have flow variations. For this reason, supporting transient CFD models is a critical issue to take into account. In this regard, HydroSludge includes tools for filtering data of transient influents flow monitored at the WRRFs. This utility allows the user to apply moving averages as well as outliers smoothing functions to the data for afterward usage during the CFD simulation.
\item{Rheology}
CFD simulation for wastewater treatment can consider non-Newtonian fluid models to represent the activated sludge i.e., Shear-thinning Model ~\citep{Samstag2016}. HydroSludge includes the Bingham, Ostwald de Waele (Power Law), and Hershel-Bulkley models to consider the fluid viscosity properties.
Users can simulate the behavior of the fluid transport model of an SST considering two options:
\begin{itemize}
\item The rheological model of sludge for non-Newtonian fluids. By default, the viscosity is modeled with the Bingham plastic model.
\item  Water as a Newtonian fluid. If this option is used, the calculation of the fluid characteristics is performed internally in HydroSludge.
\end{itemize}
Users can simulate the behavior of the fluid transport of a bioreactor considering two non-Newtonian fluid models (Ostwald de Waele, or Hershel-Bulkley) or water as a Newtonian fluid.
\end{itemize}
%\end{enumerate}
\subsubsection{Checking}\label{subsubsec:check}
After filtering, data can be checked two-fold:
\begin{itemize}
\item{SST Design}
This procedure includes the different recommendations found in the most relevant bibliography: \citep{cedex}, \citep{Tchobanoglous2003}, and \citep{Dapelo2018}); to provide to the user a straightforward mechanism to check the dimensions of the SST and its components, such as scraper or input well.
\item{State Point and Ten Layers Models}
One-dimensional models are still useful techniques for a preliminary evaluation~\citep{Ekama1997}. 
Concretely, users can check the SST performance according to 1D models. Accordingly, if the SST working conditions failed at these models, the user may check these conditions before proceeding to the CFD model development.
HydroSludge implements the two most widespread models to reproduce the SST performance. On the one hand, the State Point Model allows the global performance of the operation to be calculated through a mass balance: the mass of solids entering and exiting the tank considering the external recycling rate~\citep{Ekama1997}. From this, the optimum design return activated sludge recycle rate, at which SST is in a critically loaded condition corresponding to a stable steady state sludge blanket level, and control strategy are most accurately determined. On the other hand, the Ten Layers Model~\citep{Kynch1952} shows the distribution of the total suspended solids inside the SST. It allows the sludge blanket height and total suspended solids distribution to be calculated under specific settling velocity conditions.
\end{itemize}

\subsection{Simulation} \label{subsec:simulation}
This section details the differentiating features of HydroSludge for assisting the user in developing a WRRF CFD model with minimum effort.

\subsubsection{Submersible Mixers}\label{subsubsec:submersible}
Submersible mixers, flow boosters, or impellers are a type of equipment commonly found in WRRFs. They enhance mixing, prevent the formation of dead volumes, or just direct the flow in a specific direction. HydroSludge permits to include impellers in the CFD simulations by using the so-called equivalent momentum source~\citep{Climent2019}. In this approach, the complex impeller geometry is substituted by a cylindrical region that applies a momentum source that mimics the hydrodynamic effects of impellers at a reduced computational cost. Note that including the full impeller geometry would result in a huge increase in the number of nodes to resolve the complex geometry of the blades and their corresponding boundary layers. Furthermore, the rotation of the blades requires highly demanding computational techniques such as deformable meshing or the frozen rotor approach. 
\par
First, the volumetric momentum source term, $S_m$, is introduced into the RANS equations via the \texttt{swak4Foam} library,
\begin{small}
\begin{equation}
    S_m=M_p\ (\hat{u}_{C,x},\hat{u}_{C,y},\hat{u}_{C,z}),
\end{equation}
\end{small}
being $M_p$ the volumetric momentum modulus, and $\hat{u}_{C,i}$ the director vector components in Cartesian coordinates ($i=x,y,z$) for the impeller axis. The modulus for the volumetric momentum source is given by   
\begin{small}
\begin{equation}
    M_p=\frac{\rho}{V_M}\left(\frac{q}{D_b}\right)^2,
\end{equation}
\end{small}
where $D_b$ is the actual diameter of the blades, $\rho$ the fluid density and $V_M$ the volume of the source region. The propelled flow rate, $q$, can be calculated as
\begin{small}
\begin{equation}
    q=D_b\sqrt{\frac{\omega}{\omega_o}\frac{F_o}{\rho}}.
\end{equation}
\end{small}
In this equation, $F_o$ represents the thrust force for the design rotational speed $\omega_o$ of the propeller. These two parameters are usually included in the technical specifications supplied by propeller manufacturers. In some applications, the actual rotational speed, $\omega$, differs from the design one, so the effective flow rate must be modified accordingly. 
Second, the application of this term is limited within a cylindrical region, the so-called source region, which surrounds the actual propeller. The axis of this cylindrical region is coincident with the actual propeller axis, its diameter equals the diameter of the blade, and its center is located at the rotation center for the blades. The length of the cylinder can be set at will, but it usually spans about one-fifth of the blade diameter to permit an adequate meshing of this region.
Figure~\ref{fig:raul} illustrates the basic geometry for the cylindrical propeller subvolume. The center of the cylinder is given by the coordinates vector
${\vec{r}}_C=\left(x_C,y_C,z_C\right)$, and its axis orientation by the unit vector ${\hat{u}}_C$. The radius and axial length of the cylinder are represented by $R$ and $L$ respectively. To check if a given point P, with coordinates ${\vec{r}}_P=\left(x_P,y_P,z_P\right)$, is included within the momentum source subvolume, one must verify the following two conditions:
\begin{figure}[htp]
    \centering
    \includegraphics[trim={0 0 0 0}, clip, width=0.75\linewidth]{"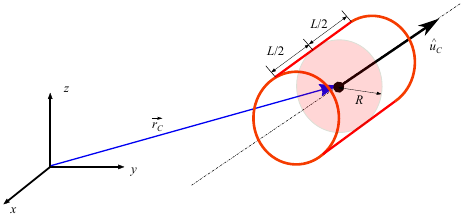"}
    \caption{Coordinate system of the momentum source volumetric region.}
    \label{fig:raul}
\end{figure}
\begin{enumerate}[(a)] % (a), (b), (c), ...
\item 	The distance between the point and the cylinder axis, $d_{Pu}$, must be smaller or equal than the radius. This distance can be calculated as
\begin{small}
\begin{equation}
d_{Pu}=||(\vec{r}_p - \vec{r}_c)\hat{u}_C||.
\end{equation}
\end{small}
\item 	Point $C$ and vector $\hat{u}_C$ define a plane, $\pi_C$. The distance between this plane and the point $P$, $d_{P\pi}$, must be smaller or equal than $L/2$, being:
\begin{small}
\begin{equation}
   d_{P\pi}=\left|u_{C,x}\left(x_P-x_C\right)+u_{C,y}\left(y_P-y_C\right)+u_{C,z}\left(z_P-z_C\right)\right|.
\end{equation}
\end{small}
\end{enumerate}

\subsubsection{Surface Selection and Meshing}\label{subsubsec:surface}
HydroSludge provides a user interface to select and determine surfaces and boundaries from CAD geometries.
The surfaces and boundaries are sets of faces that users directly define in the geometry. These objects are populated to the meshing tool, and ultimately to the CFD simulation.
These surfaces are automatically defined in the meshing process to proceed to their fitting and refinement.
This geometry-to-mesh tool is one of the cornerstones of the HydroSludge effort to facilitate the work to its users.
Once the surfaces and boundaries are named and configured, HydroSlduge divides the meshing into four stages:
%This feature of the software allows the user to import any geometry and not just WRRF components. For this reason, this stage establishes the application end-point depicted in Figure~\ref{fig:dataworkflow}.
\begin{itemize}
\item Blocking: Cube-shaped hexahedral mesh generation around the imported geometry. Hexahedral cells can be adjusted in size to achieve more or less mesh refinement. A finer mesh implies more accurate results, but also a higher computation time in both meshing and simulation processes.
%Conversely, a thicker mesh provides less exact results but lower calculation times. In this regard, a balance must always be sought in the refinement of the mesh.
\item Sculpting: Refinement of cells, vertices, and edges. This operation, employing an iterative process, divides the cells into hexahedrons. At this point, a first hexahedral approximation of the geometry is ready.
\item Snapping: Projection of cells vertices on the surface of the geometry to removing areas with sharp edges. In other words, smoothing pronounced edges in the geometry.
\item Layering: Addition of extra layers for refining and improving the quality of the hexahedral mesh, especially on the outer faces of the geometry. HydroSludge adds a user-defined number of layers with an expansion ratio in the declared surfaces.
\end{itemize}

\review{
Finally, the OpenFOAM \texttt{checkMesh} tool is also included in HydroSludge, allowing users to know the quality of the meshes.
This tool returns, as a result, the total number of cells, points, and hexahedrons of a mesh, as well as the quality of the mesh in terms of aspect ratio and skewness metrics. 
%In the Aspect Ratio, the parameter used to calculate the quality of the mesh is based on the ratio of the longest edge to the shortest edge for every cell. Ideally, it should be close to 1. In Skewness, the parameter that is used to calculate the quality of the mesh is calculated for every face connecting neighboring cells. A point in this face is defined as the intersection between the line joining cell centers and the face itself. Then, the ratio of the distance between this point and the face centroid to the distance between cell centers is calculated. Ideally, it should approach 0 indicating that the intersections occur at the face centroids.
}

\subsubsection{Solvers}\label{subsubsec:solvers}
HydroSludge leverages OpenFOAM features, and it includes three different WRRF-related solvers.
\paragraph{pimpleFoam} Large timestep transient solver for incompressible turbulent single-phase flow. This solver is based on the PIMPLE algorithm which is a combination of:
pressure implicit split operator (PISO), used in transient problems; and semi-implicit method for pressure-linked equations (SIMPLE), used in steady-state simulations.
These iterative algorithms solve equations for velocity and pressure.
\paragraph{driftFluxFoam} Relative motion between two incompressible phases solver. This solver is mainly used in SSTs which have a continuous phase and a solid dispersed phase. Transient analysis can be performed, introducing a dynamic influent flow to calculate the evolution of the Sludge Blanket Height (SBH) and the Sludge Concentration Distribution (SCD) within the SST. 
%A non-newtonian flow with complex rheological model can be considered.
\paragraph{ASM1Foam} Biochemical behavior model after obtaining the hydrodynamic results. ASM1~\citep{henze1987} model is included in HydroSludge through an extension of OpenFOAM solver \textit{scalarTransportFoam}. At this solver, the state variables are introduced as $k$ scalars with the turbulent transport equation:
\begin{small}
\begin{equation}
\frac{\partial\phi_k}{\partial t}+\nabla(U\phi_k\ )-\nabla^2\bigg[\bigg(D_T+\frac{\nu_t}{Sch_k}\bigg)\phi_k\bigg]=S_k
\label{eq:ros1}
\end{equation}
\end{small}
being $\phi_k$ the concentration of the $k$ scalar, $U$ the fluid velocity, $D_T$ the molecular diffusion coefficient divided by the fluid density, $\nu_t$ the kinematic viscosity, $Sch$ the Schmidt Number and $S_k$ the source term for $k$ state variable of the biological model. 

\section{Software Features}\label{sec:software}
This section thoroughly depicts the software design, library dependencies, and implementation details to tackle the previously presented methods.
The section ends with a description of the usage modes provided by HydroSludge.

\subsection{Implementation}\label{subsec:implementation}
\review{
HydroSludge core is written in C++ and is empowered by the
following widely used open source libraries with highly active communities\footnote{\url{https://dev.opencascade.org/forums}}\footnote{\url{https://discourse.vtk.org}}\footnote{\url{https://forum.qt.io}}:
}

\begin{itemize}
\item Qt widget toolkit\footnote{\url{https:// www.qt.io}}  for the GUI (Graphical User Interface).
\item OpenCascade Technology (OCCT) software development platform\footnote{\url{https://dev.opencascade.org}}  for 3D computer-aided design (CAD). 
\item Visualization Toolkit (VTK) software system\footnote{\url{https://www.vtk.org}} for 3D computer graphics and visualization.
\end{itemize}
\par
Furthermore, HydroSludge includes a virtualized subsystem that leverages Docker for executing OpenFoam operations. Docker uses an operating system (OS) level virtualization through packages named containers. Within these containers, a user can deploy customized instances of an OS with the necessary dependencies. This container runs with complete independence and in isolation on the host OS.
In this regard, HydroSludge installation binaries include a Docker image with the software packages required in order to provide the necessary tools to perform all the operations brought to the users by the GUI. 
Concretely, The container released with HydroSludge is based on Ubuntu\footnote{\url{https://ubuntu.com/containers}} 18.04 LTS and satisfies the dependencies for running the following software, also packed in the image:
\begin{itemize}
\item OpenFoam\footnote{\url{https://openfoam.com}} v19.06 
\item Swak4Foam\footnote{\url{http://openfoamwiki.net/index.php/Contrib/swak4Foam}}
\item MPICH\footnote{\url{https://www.mpich.org}} 3.3.2
\end{itemize}
\par
\begin{figure}[htp]
    \centering
    \includegraphics[trim={1.25cm 0.15cm 0.175cm 3.7cm}, clip, width=0.7\linewidth]{"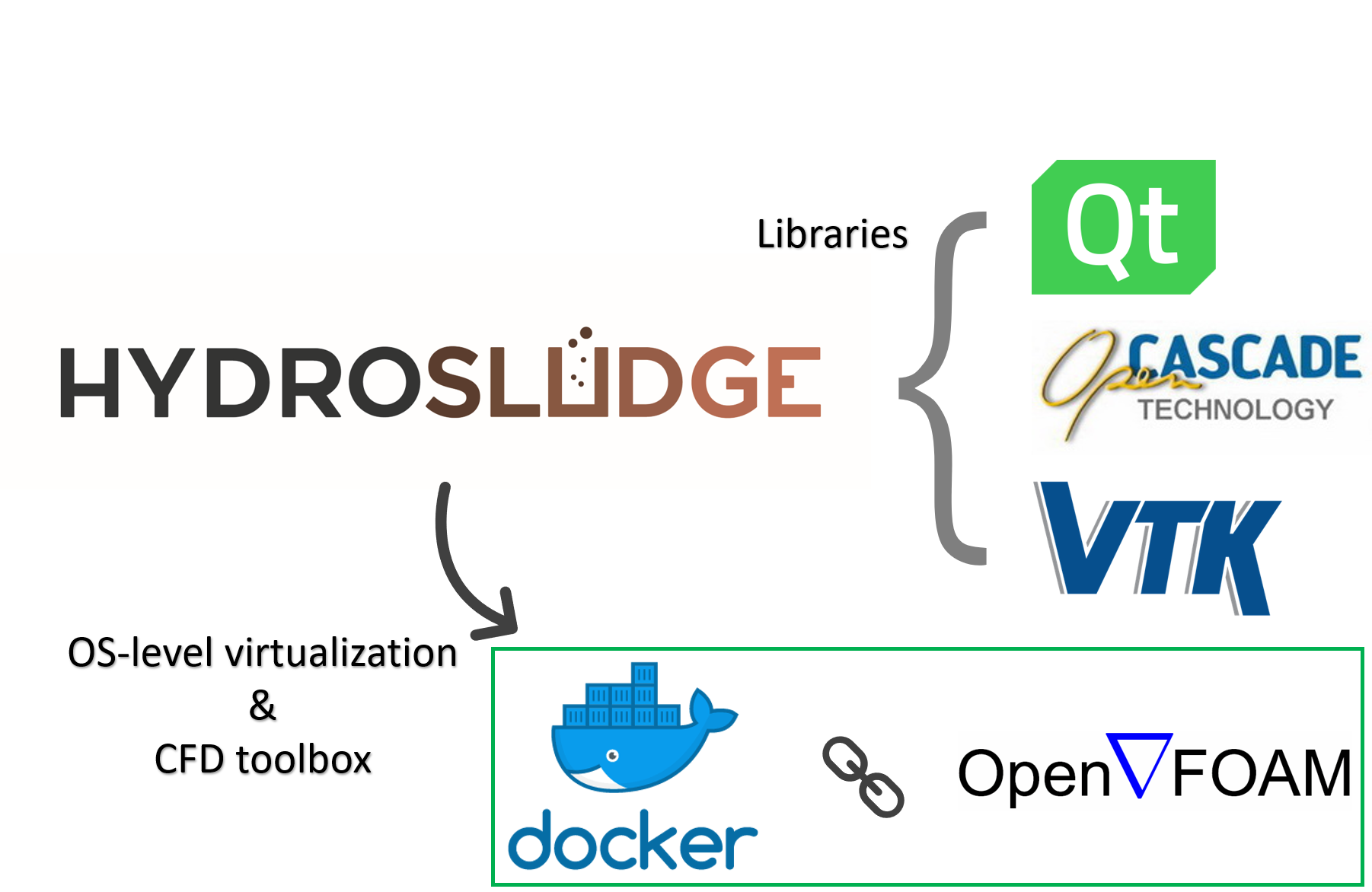"}
    \caption{Overview of third-party software integration in HydroSludge.}
    \label{fig:hs-overview}
\end{figure}

Summarizing, Figure~\ref{fig:hs-overview} lists the libraries leveraged by HydroSludge, and depicts the virtualized environment for running the simulations.
\par
\begin{figure}[htp]
    \centering
    \includegraphics[width=0.8\linewidth]{"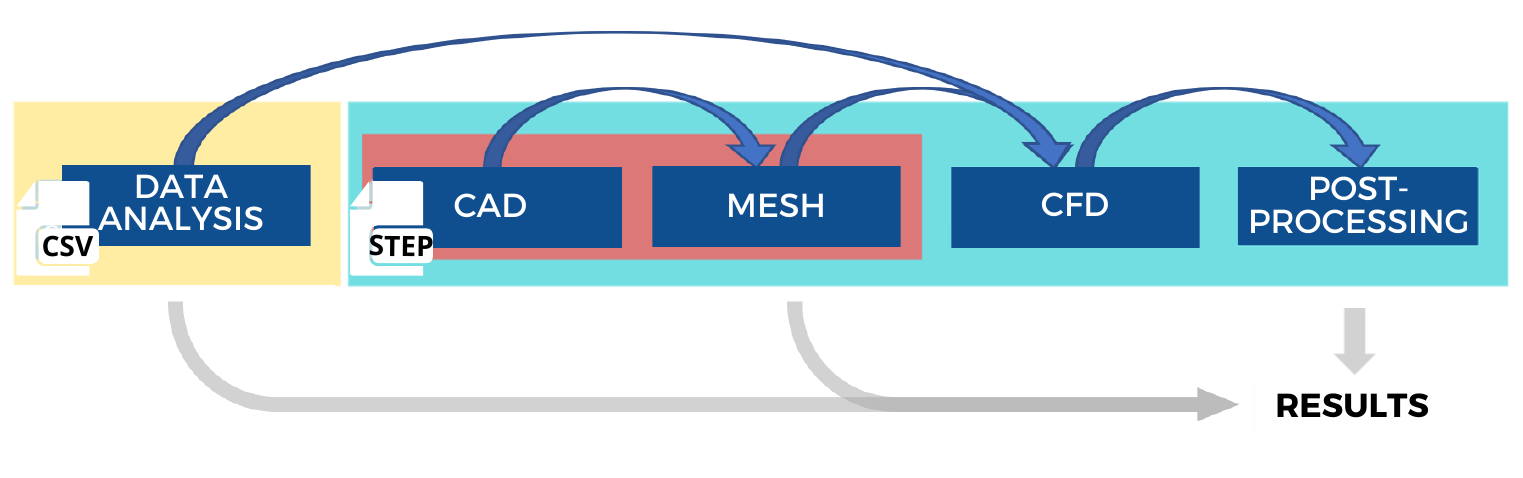"}
    \caption{HydroSludge modules and operations.}
    \label{fig:dataworkflow}
\end{figure}

HydroSludge GUI modular design is represented in Figure~\ref{fig:dataworkflow}.

In the figure, five modules can be appreciated. Following they are described.

\subsubsection{Data Analysis}
\review{
This module implements two methods for processing the input data described in Section~\ref{subsec:preliminary}.
\newline\indent
The first method, ``outliers-removal'', scans the input data selecting small subsets of n elements. The algorithm continues with the calculation of the mean and standard deviation for the subset data. Then, every element in the subset is compared to the subset mean and replaced by it if the element lies outside the $1.96$ standard deviation range. 
\newline\indent
The second method, ``moving-average'', scans every element in the dataset and replaces it with the mean value of the $n$ closest neighbors. 
\newline\indent
Leveraging VTK versatility, users are capable of preparing the data, configuring, and visualizing a series of charts included for this purpose. In this regard, HydroSludge makes use of the VTK filtering methods for reshaping and fitting the inflow. Likewise, it also provides adjustment mechanisms for calculating the rheological properties necessary for the afterward simulation.
}

\subsubsection{CAD}\label{subsubsec:cad}
HydroSludge explores an imported STEP-File\footnote{\url{https://webstore.ansi.org/Standards/ISO/ISO10303212016}} with the shape of the object to mesh. Leveraging OCCT, each face is identified and labeled in its center of mass. In this regard, a user is enabled to select faces and group them to determine areas. These areas are crucial to define boundary conditions in further steps. Furthermore, at this point, areas are devised as \textit{patch} or \textit{wall} depending on their role in the simulation. Surfaces are automatically meshed following the user’s guidelines of areas and specifications.

The linking process between the geometry and the mesh is the creation of a series of \texttt{.stl} files corresponding to the surfaces and boundaries defined by the user. In this process, the initial mesh is built from the associated shapes of the geometry concerning their correctly triangulated parts. This operation is based on the OCCT method: \texttt{RetrieveMesh}.
The \texttt{.stl} files provide OpenFOAM the flexibility of applying mesh refinement methods to a specific boundary.
This underlying operation is crucial for a satisfactory and productive user experience. For this reason, HydroSludge acts as a black box where users are only responsible to determine the level of refinement in each surface via a series of input fields in the GUI.

\subsubsection{Mesh}
CFD simulations rely on meshes. OpenFOAM provides tools for defining and refining a mesh, which is a sequential process that HydroSludge handles. For this purpose, a critical aspect is that OpenFOAM needs to load the geometry in a supported format, such as STL\footnote{\url{https://www.loc.gov/preservation/digital/formats/fdd/fdd000506}}, which describe an unstructured triangulated surface by unit normal and vertices in a 3D Cartesian system. 
For this reason, the HydroSludge meshing process internally includes the tailoring of STL files ready for OpenFOAM. 
As previously introduced (see Section~\ref{subsec:simulation}), OpenFOAM meshing involves a series of iterative steps so that, HydroSludge provides an interface divided into four stages: 
\begin{inparaenum}[i)]
\item blocking,
\item sculpting,
\item snapping, and
\item layering.
\end{inparaenum}
\par
HydroSludge, in the first stage leverages the OpenFOAM command \texttt{blockMesh} in order to perform the \textit{blocking}. This command reads the configuration from the dictionary file \textit{blockMeshDict}. In this operation, OpenFOAM registers surfaces’ names and types previously defined by the user from the geometry faces.
\par
Once the initial mesh has been generated with \texttt{blockMesh}, OpenFOAM provides the command \texttt{snappyHexMesh} in order to perform the remaining stages. \texttt{snappyHexMesh} reads the execution parameters from \textit{snappyHexMeshDict} file, which depending on its configuration can process the \textit{sculpting}, the \textit{snapping}, or the \textit{layering}.
However, the user will remain agnostic to the OpenFOAM technical details. 
\par
The sequential nature of the meshing process allows the user to refine each stage, or come back to a previous stage until reaching a satisfactory result.
Finally, users can check the mesh quality by clicking a button that executes the OpenFOAM command \texttt{checkMesh}.
During the whole process, HydroSludge leverages the VTK library to visualize the results in a viewer.

\review{HydroSludge also provides an \textit{Advanced Control} option to further fine-tune the meshing parameters. This feature enables the direct edition of the OpenFOAM configuration file.
Advanced users, with high experience in modeling, may make the most of this option to increase the quality of the resultant mesh.}

\subsubsection{CFD}\label{subsubsec:cfd}
Once the mesh is ready, the CFD module assists the user to configure the solver parameters, (importing them from the data analysis stage, if needed). Furthermore, the boundaries defined in the geometry are presented accordingly to the selected solver characteristics. 
Depending on the solver or equipment included in the case, HydroSludge handles internally the necessary files (i.e.:  \textit{fvOptions}, \textit{fvSchemes}, \textit{fvSolution}, \textit{transportProperties}...).
Finally, the rest of the control simulation parameters are also configured.
\par
HydroSludge leverages OpenFOAM MPI multiprocess capabilities to accelerate the simulation in multicore environments. Concretely, HydroSludge decomposes the domain in a given number of processes (up to the number of CPUs assigned to Docker) and defines a processor topology by editing the OpenFOAM file \textit{decomposeParDict} and running the command \texttt{decomposePar}. Notice that parallel executions are also available for the meshing process.
\par

\subsubsection{Post-processing}
A series of predefined operations are included for the post-processing stage. HydroSludge provides tools for defining planes, creating isovolumes, contouring surfaces, and plot probe lines through the object.
This set of operations are based on VTK methods, besides they have been tailored to the most common WRRF operator necessities and present a reduced configuration complexity.

%If users required more advanced or specific utilities, an instance of Paraview with the current case loaded could be launched from HydroSludge (Paraview needs to be installed in the host machine).

Every single OpenFOAM utility (either launched with MPI or not) via the command \texttt{docker run} via the Windows \texttt{PowerShell}.
HydroSludge enables real-time visualization of the simulation results using the VTK library. In this regard, HydroSludge reads on-the-fly the timesteps and provides a series of simplified tools for scientific visualization of structures.
Notice that these features are also compatible with parallel execution. In this case, HydroSludge internally performs the reconstruction operation for each new solved timestep with the OpenFOAM command \texttt{reconstructPar}, completely transparent to the user.

\begin{figure}[htp]
    \centering
    \includegraphics[width=0.8\linewidth]{"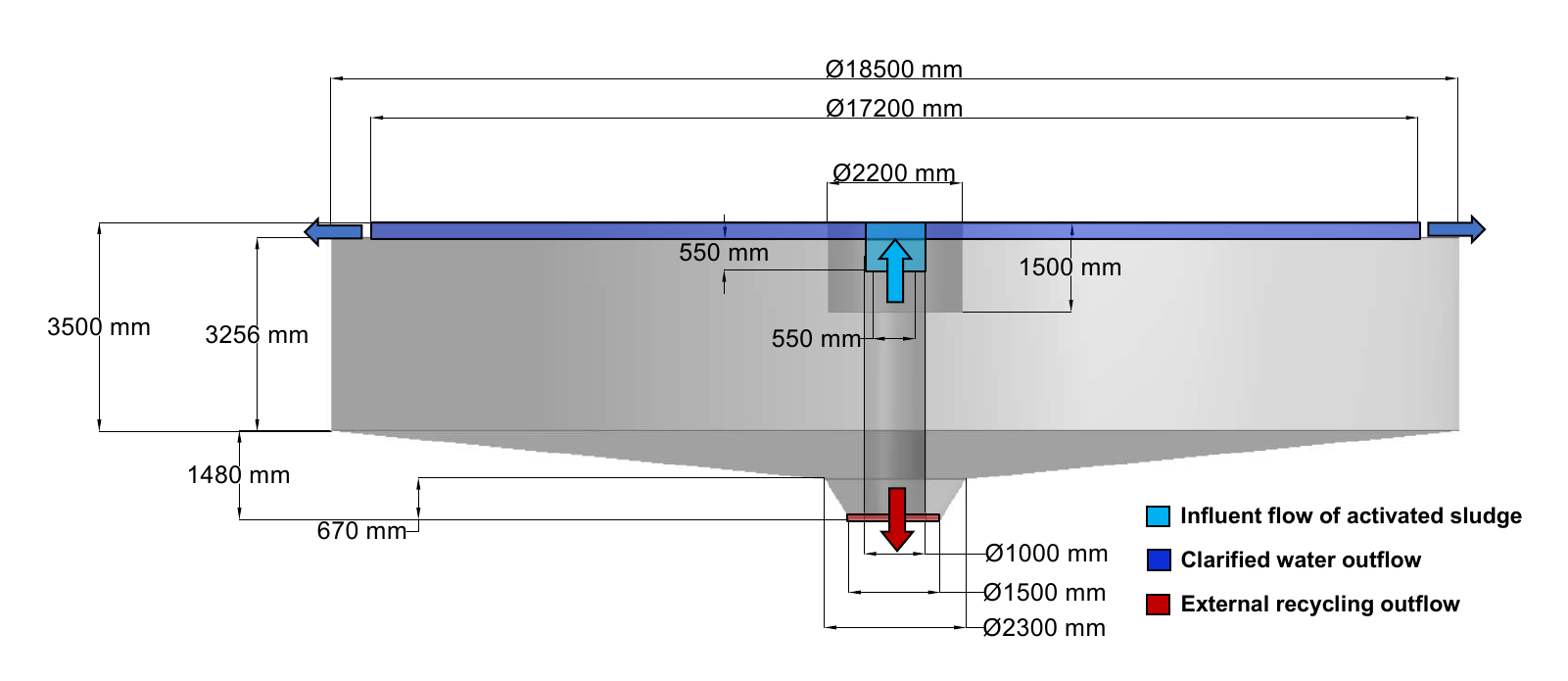"}
    \caption{Scheme of the in/out flows in the structure of the SST under study.}
    \label{fig:cotas}
\end{figure}

\subsection{Usability}\label{subsec:usability}
HydroSludge design principles arose from the need to provide useful and efficient management from a facility technician to the manager of the WRRF.
In this regard, it is not mandatory to make comprehensive use of all the features provided by HydroSludge. Instead, HydroSludge allows flexible usage by providing different entry and output points. 
Figure~\ref{fig:dataworkflow}, apart from the modules priorly described (blue boxes), also depicts HydroSludge input points (white files), data dependencies among modules (blue arrows), and intermediate or final results (gray arrows pointing to ``Results''). 
\par
Although HydroSludge is conceived as a single tool, its modular design allows it to use it as a toolbox of connected components. For this purpose, HydroSludge also includes independent procedures that lead to different types of results. Figure~\ref{fig:dataworkflow} depicts with primary-colored squares the three operations that can be independently performed.
\begin{enumerate}
\item Users are capable of conducting a preliminary study through the \textit{Data analysis} module (yellow square). Left to right, Figure~\ref{fig:dataworkflow} shows this first module, which independently can obtain results, and it expects a series of \texttt{.csv} files with data of the flow and the settling performance. Particularly, this module provides tools for exploring the appropriateness of the characteristics of the SST physical design, the transient inflow data, the settling models, the rheology of the target case, and finally, how the influent and the external recycling flow rates affect the SST performance. These methods are thoroughly described in Section~\ref{subsec:preliminary}.
\item HydroSludge can be used as a standalone tool for generating a mesh from CAD geometries. The red square in Figure~\ref{fig:dataworkflow} highlights the modules involved in the meshing operation. By importing a \texttt{step} geometry in the \textit{CAD} module, users are provided with the necessary instruments to create and refine a mesh in the \textit{MESH} module.
%Section~\ref{subsubsec:surface} described its usage.

\item From a holistic point of view, HydroSludge comprehends a complete CFD simulation. The blue square in Figure~\ref{fig:dataworkflow} groups the necessary modules for configuring and running a simulation (\textit{CFD} module), and its visualization (\textit{Post-processing} module). Furthermore, optionally, \textit{Data analysis} module can also take part in the simulation process.
\end{enumerate}

\section{Results}\label{sec:results}
In this section, two study cases are presented and developed. The study of these cases serves as a proof of concept and correct operation of HydroSludge.

\subsection{Study of a Secondary Settling Tank}\label{subsec:SST}
The full-scale SST is 18.5m in diameter and 5m in height.
%The first study case is an SST located at a WRRF in Tineo (Spain). The WRRF only has one SST in operation, for this reason, all activated sludge of the facility passes through this unit process. With more detail, the studied unit is 18.5 meters wide, and more than five meters high. 
With more detail, Figure~\ref{fig:cotas} depicts the exact dimensions of the studied SST. 
The figure also indicates the location of the inlet, which is the influent flow of activated sludge; and the two outlet flows, one for clarified water, and the other for the external recycling.

\subsubsection{Data Analysis}
The study of this case starts with a preliminary study leveraging the tools included in the data analysis module  (see Section~\ref{subsec:preliminary}).

\begin{figure*}[htp]
    \centering
    \begin{subfigure}[t]{0.5\textwidth}
        \centering
        \includegraphics[trim={2.5cm, 1cm, 6cm, 1.25cm}, clip, width=\linewidth]{"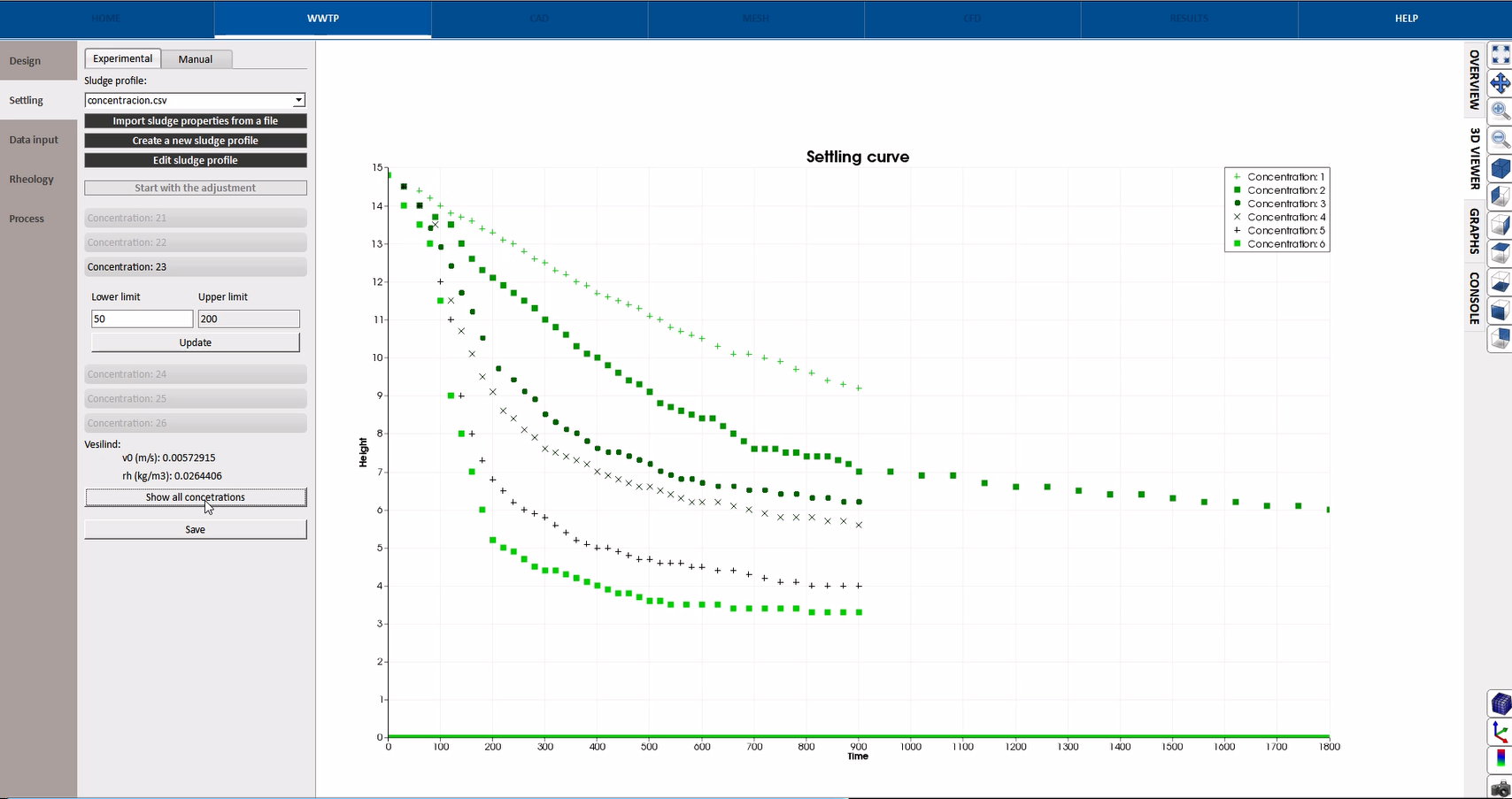"}
        \caption{Linear adjustment of the settling.}
        \label{subfig:sett-meas}
    \end{subfigure}%
    ~ 
    \begin{subfigure}[t]{0.5\textwidth}
        \centering
        \includegraphics[trim={2.5cm, 1cm, 6cm, 1.25cm}, clip, width=\linewidth]{"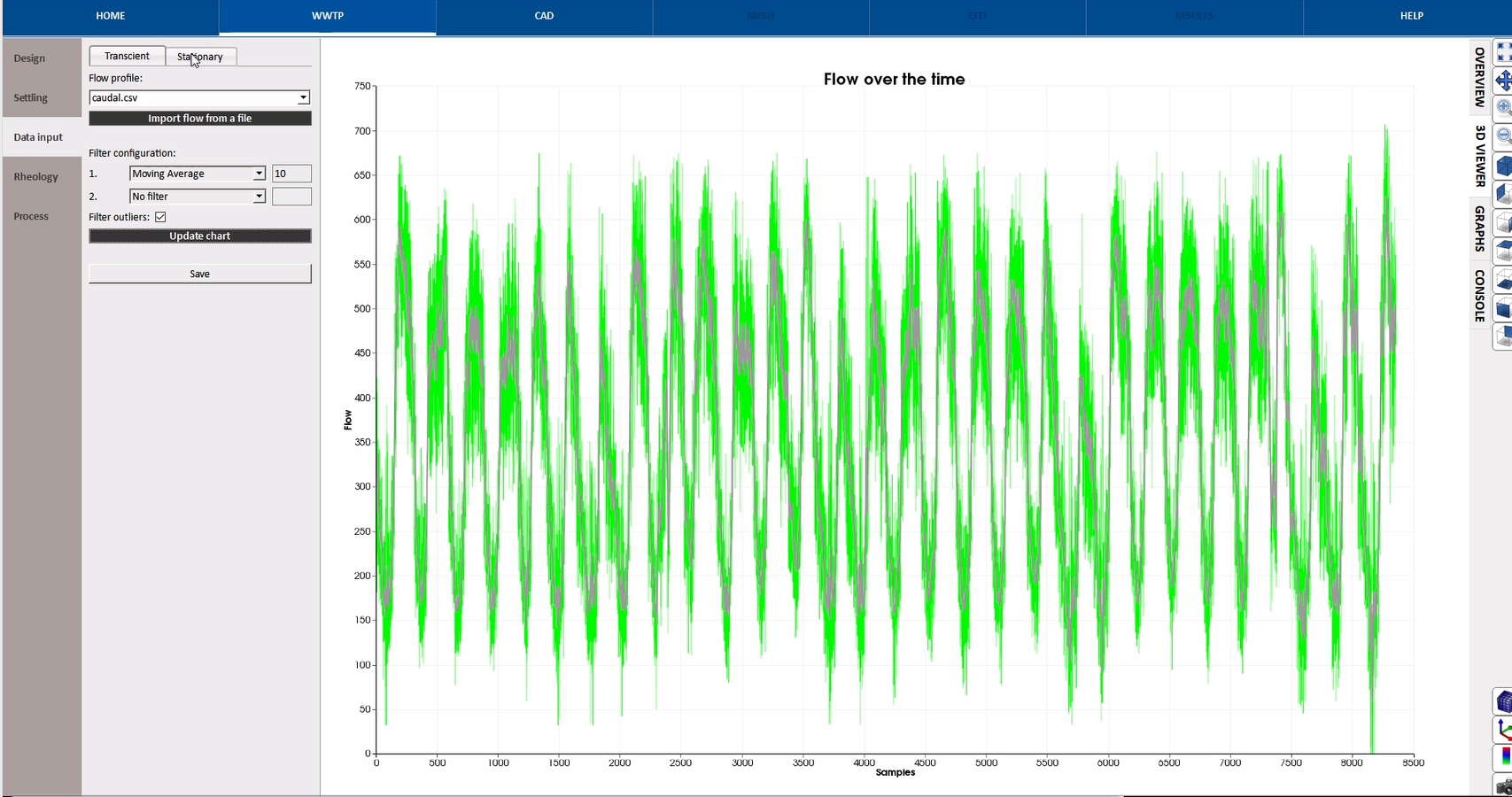"}
        \caption{Filters over the inflow.}
        \label{subfig:inflow}
    \end{subfigure}
    \caption{Examples of the filtering stage of the SST design tool.}
    \label{fig:Process}
\end{figure*}

\paragraph{Filtering}
With the \texttt{Settling} tool, users calibrate the settling model parameters. For this purpose, firstly, experimental data of the batch settling test are introduced importing a \texttt{.csv} file or manually.
Next, the linear equation is fitted to the linear part of the experimental settling curve for each activated sludge concentration. When all linear curves are fit to the experimental results, HydroSludge calculates the settling model parameters (see Figure ~\ref{subfig:sett-meas}). This procedure is not mandatory so that, in this case, the settling model is introduced manually.
\par
The \texttt{Inflow} tool, allows users to obtain a smooth transient flow from monitored transient influent flows. Firstly, a \texttt{.csv} file with the evolution data of the flow through the time is imported. Users are provided with a series of tools to apply filters, such as moving averages or outlier removal (see Figure~\ref{subfig:inflow}). The filtered transient flow can be used at the CFD simulation afterward and exported. 

Finally, the \texttt{Rheology} tool saves the rheological model selected for this case and transfers the configuration to the CFD module, if needed.
%The Bingham model was chosen for the characterization of the rheological behavior of the sludge in this study.

\paragraph{Checking}
With the \texttt{Design} tool users check if the dimensions of the SST agree with bibliography design requirements.
\par
\begin{figure*}[htp]
    \centering
    \begin{subfigure}[t]{0.5\textwidth}
        \centering
        \includegraphics[trim={0cm 9cm 1.5cm 1.25cm}, clip, width=\linewidth]{"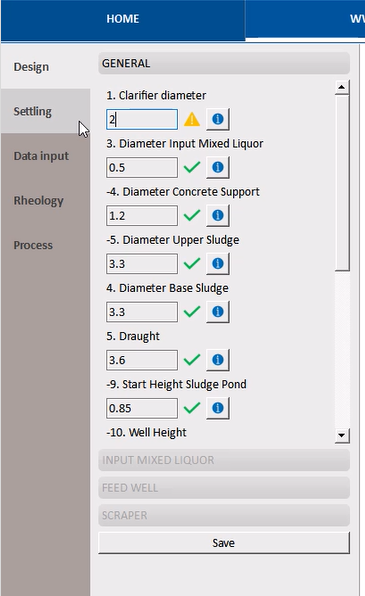"}
        \caption{Structural design configuration.}
            \label{subfig:warnings}
    \end{subfigure}%
    ~
    \begin{subfigure}[t]{0.5\textwidth}
        \centering
        \includegraphics[trim={9.6cm 0cm 3cm 2cm}, clip, width=\linewidth]{"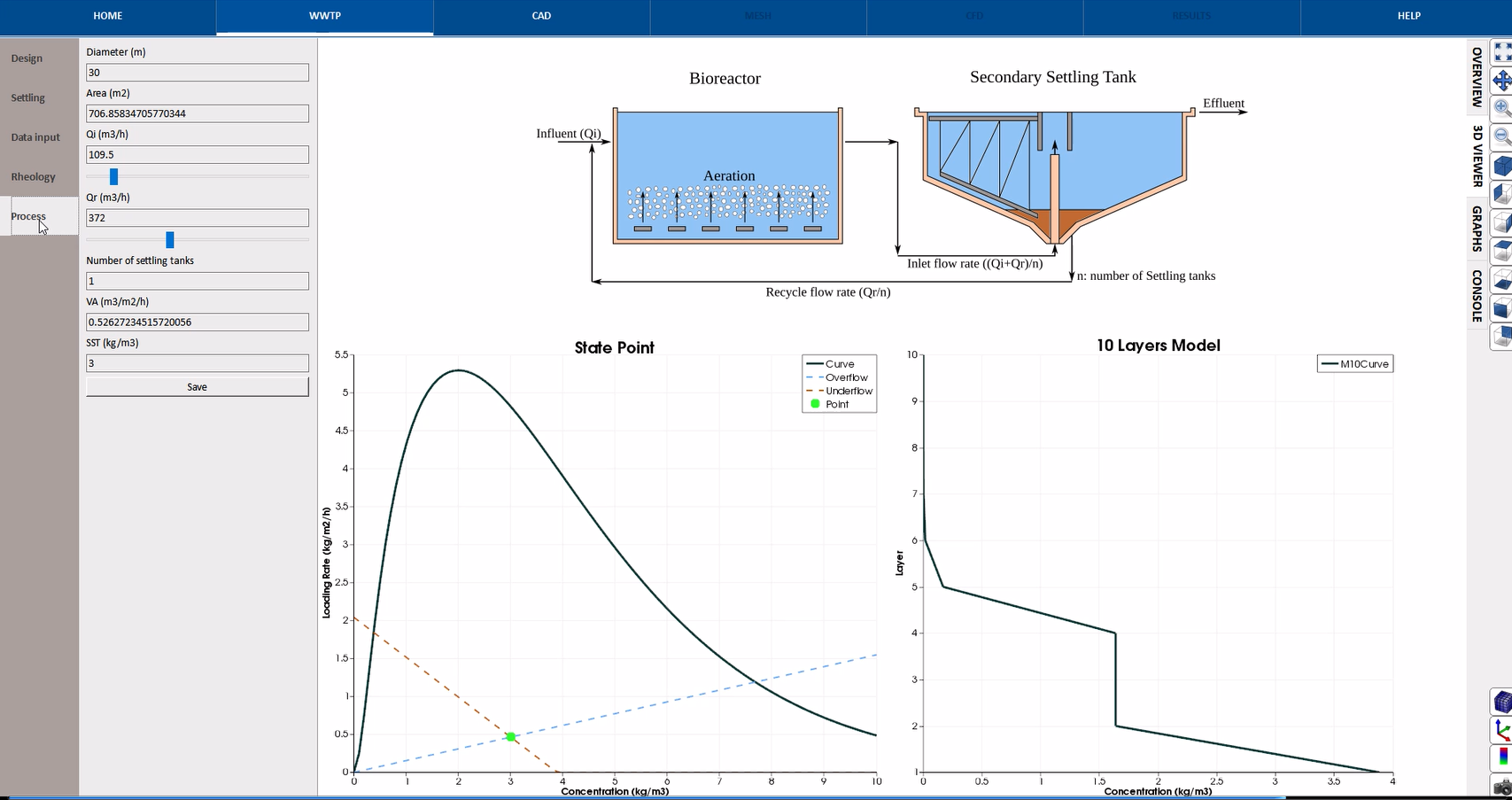"}
        \caption{Sludge settling process.}
        \label{subfig:ten-ok}
    \end{subfigure}%
 
    \caption{Checking stage of the SST design tool.}
    \label{fig:SSTDesign}
\end{figure*}

HydroSludge indicates with a check-mark or a warning sign whether the dimension is acceptable (see Figure~\ref{subfig:warnings}). More information can be obtained clicking on the information button in order to get a dialogue with recommended design information and their reference.
\par
All in all, the filters and the previous check can be used by the state point model and the ten layers model to test whether the activated sludge from the biological reactor settles at the SST.
This tool is presented with three areas (see Figure ~\ref{subfig:ten-ok}). 
The first area, found at the top, is a hydraulic diagram of the secondary treatment of a WRRF with a bioreactor and an SST. 
The second area at the bottom corresponds to the 1D models to represent the SST performance.
\par
The models are configured using the data introduced, such as the diameter and area of the SST, the influent and recirculated flow (Qi and Qr, respectively), number of settling tanks, overflow velocity (VA), and the activated sludge concentration (TSS). According to those parameters, HydroSludge calculates the underflow and overflow lines and the Ten Layers Model. 
The State Point of the WRRF is the underflow-overflow intersection. 
In this regard, the user can modify and explore the limits of its State Point depending on the given parameters. 
Thus, the boundary conditions can be checked before proceeding with the CFD model.
%Therefore, if the State Point is inside the obtained curve, the settlement of the sludge at the SST is correct, and the Ten Layers Model behaves as expected (see Figure~\ref{subfig:ten-ok}). 
%In contrast, if the point is outside the curve, the sludge will not settle properly at the SST, and the solids concentration will remain constant at the ten layers model. In this case, flow rates, dimensions, and sludge concentration of the SST should be adjusted before proceeding to the CFD model.

\review{
The design requirements have been extracted from the following national and international sources:~\cite{cedex, wef, Committee2014, Tchobanoglous2003}.}

\subsubsection{Boundaries Definition}\label{subsubsec:boundaries}
\begin{figure*}[htp]
    \centering
    \includegraphics[trim={11cm 2.5cm 3.25cm 1.9cm}, clip, width=0.65\linewidth]{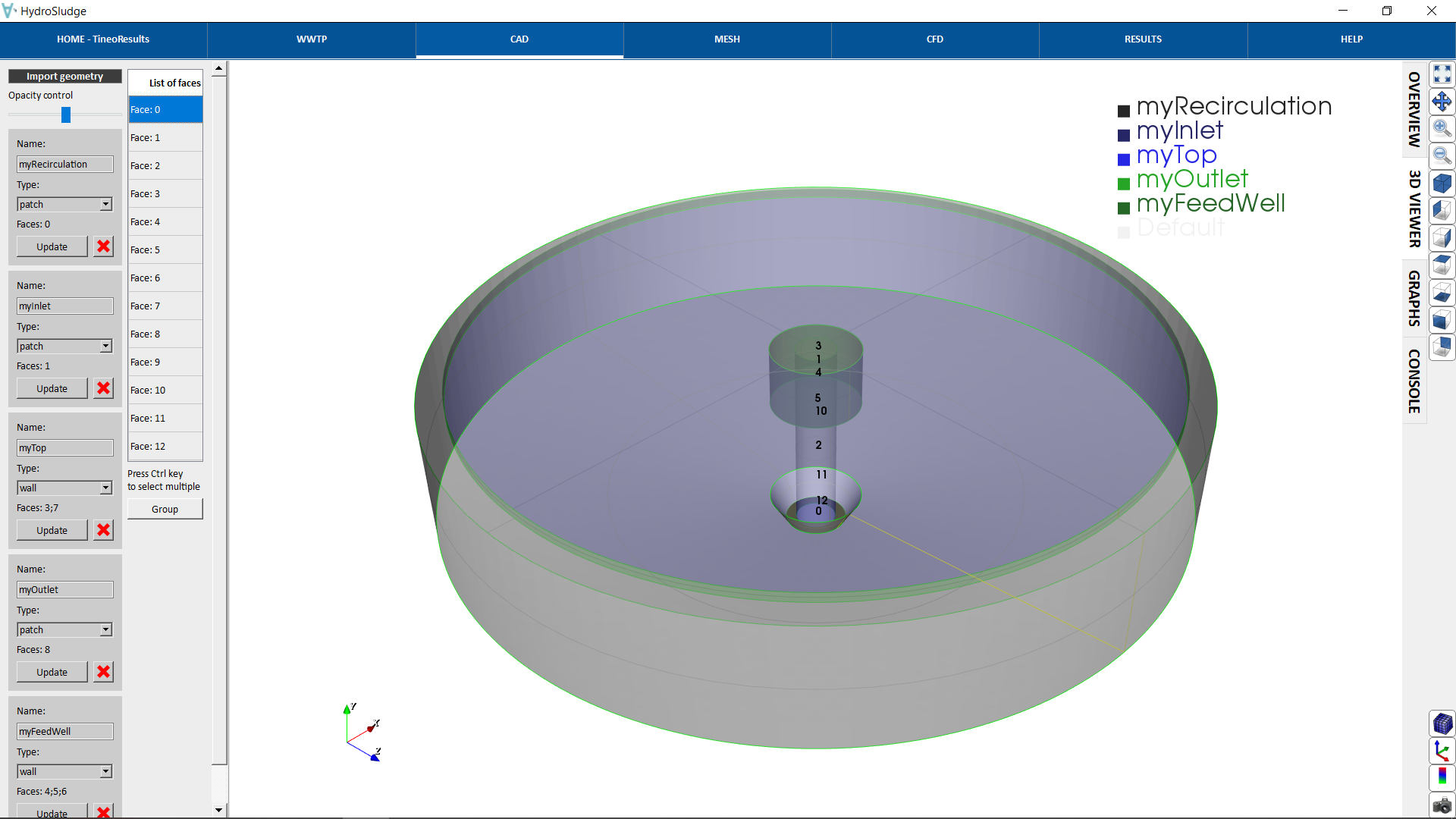}
    \caption{Surfaces of the geometry under study.}
    \label{fig:case1}
\end{figure*}

%Before starting the simulation the CFD software needs an appropriate mesh with the domain of the case. 
HydroSludge assists the complete operation from the geometry to the mesh and its refinements, as previously stated. Thus, the first step for the construction of the CFD model in HydroSludge is to import a geometry with the \textit{CAD} module. For this purpose, utilizing a \texttt{.step} file users will have the geometry faces at their disposal. Following this, users can declare the mesh surfaces and their roles in the simulation. 
For instance, the SST presented below illustrates the power of this meshing tool. Table~\ref{tab:faces} depicts an example of how faces are grouped into surfaces and these, in turn, are labeled with a name and identified as \textit{patch} or \textit{wall} according to its boundary type (see Figure~\ref{fig:case1} and Table~\ref{tab:faces}). Non-grouped faces are added to the \textit{Default} surface and identified as \textit{wall} automatically. 

\begin {table}[htp]
\small
\caption{Example of association of faces into surfaces and their types and roles as boundaries in the simulation for the study of the SST.}
\label{tab:faces}
\setlength{\tabcolsep}{5pt}
\begin{center}
 \begin{tabular}{||c c c c||} 
 \hline
Faces Id. & Surface Name & Surface Type & Boundary Type \\
 \hline\hline
 1 & myInlet & Patch & Inflow \\ 
 \hline
 0 & myRecirculation & Patch & Outflow \\ 
 \hline
 8 & myOutlet & Patch & Outflow \\ 
 \hline
 3, 7 & myTop & Wall & Atmosphere \\ 
  \hline
 4, 5, 6 & myFeedWell & Wall & Wall \\ 
 \hline
 Remaining & Default & Wall & Wall \\ 
 \hline
\end{tabular}
\end{center}
\end{table}

Thereupon, the mesh is ready to be generated in the \textit{Mesh} module. At this stage, users have access to the \texttt{Base}, \texttt{Sculpt}, \texttt{Snap}, and \texttt{Layers} meshing tools to refine the areas or surfaces of their most interest.
When the mesh is ready, it is time to configure the simulation parameters to start with the calculations of the CFD model (see Section~\ref{subsubsec:solvers}). This configuration is done at the \textit{CFD} module.

\subsubsection{Evaluation}\label{subsubsec:evaluation}

\begin{figure*}
    \centering
    \begin{subfigure}[t]{0.7\textwidth}
        \centering
        \includegraphics[width=\linewidth]{"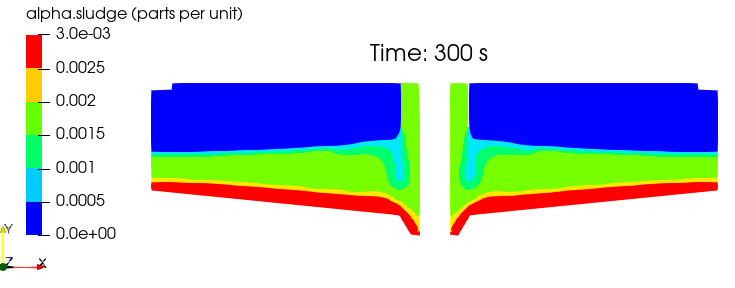"}
        \caption{HydroSludge results}
    \end{subfigure}%
    \\ 
    \begin{subfigure}[t]{\textwidth}
        \centering
        \includegraphics[width=0.6\linewidth]{"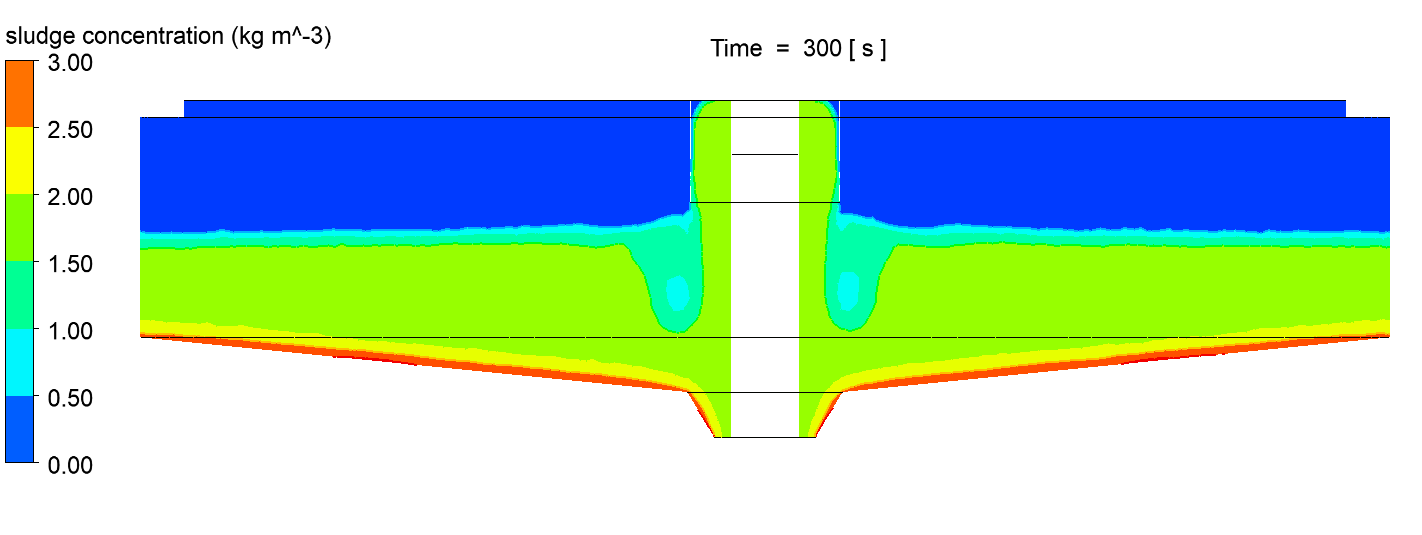"}
        \caption{Commercial CFD results}
    \end{subfigure}
    \caption{Sludge concentration inside the clarifier at the initial timesteps of the simulation.}
    \label{fig:comparisson}
\end{figure*}

\begin{figure*}
    \centering
    \begin{subfigure}[t]{0.7\textwidth}
        \centering
        \includegraphics[width=\linewidth]{"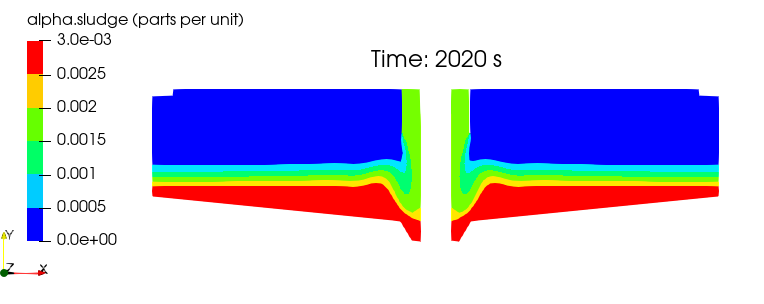"}
        \caption{HydroSludge results}
    \end{subfigure}%
    \\ 
    \begin{subfigure}[t]{\textwidth}
        \centering
        \includegraphics[width=0.6\linewidth]{"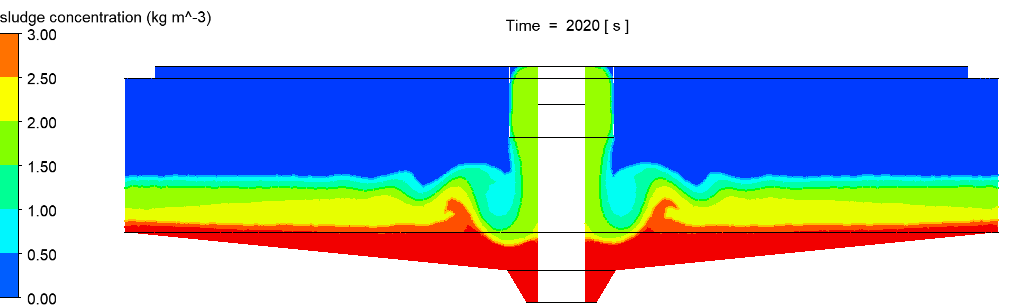"}
        \caption{Commercial CFD results}
    \end{subfigure}
    \caption{Sludge concentration inside the clarifier at 2020 seconds of the simulation.}
    \label{fig:comparisson2020}
\end{figure*}

The evaluation of the developed model in this study case is performed using \textit{driftFluxFoam} solver. The simulations start with a pre-initialized state that defines the sludge layer inside the SST. Accordingly, the user specifies the height and concentration of the sludge at the initial time of the simulation.

The last step before launch the simulation is to provide the control configuration parameters such as the number of time-steps, writing interval, or time steepness, which are defined at this stage. 

Furthermore, the parallel execution of the case can be enabled. For this purpose, the desired quantity of CPUs involved in the simulation, as well as, the processes topology, have to be introduced during the control configuration.

The results obtained by Hydrosludge are comparable to the ones calculated in a commercial CFD platform such as ANSYS\footnote{\url{https://www.ansys.com/products/fluids}}. \review{ To illustrate this, Figure~\ref{fig:comparisson} and Figure~\ref{fig:comparisson2020} depict the sludge concentration at a vertical plane of this study case obtained with HydroSludge and ANSYS 19.2. While Figure~\ref{fig:comparisson} illustrates the initial timesteps of the simulation, Figure~\ref{fig:comparisson2020} shows the results obtained when a more steady state is reached.} Both simulations show a similar sludge height blanket and similar influent flow behavior. Additionally, both simulations showed a similar sludge concentration distribution with the presence of settled solid particles at the bottom of the clarifier being evacuated through the recirculation outlet by gravity.

\subsection{Study of an Anoxic Reactor}\label{subsec:anoxic}
The second study case is a full-scale anoxic reactor from a Modified Ludzack-Ettinger biological reactor, which was extensively analyzed in~\citep{Climent2018}. Particularly, this case considers two influent flows, the influent wastewater flow, and the internal recirculation flow; the outflow, and; two submersible mixers.

\subsubsection{Equipments}
\begin{figure*}
    \centering
    \begin{subfigure}[t]{0.5\textwidth}
    \centering
    \includegraphics[trim={2.4cm 1cm 3.6cm 1.45cm}, clip,
    width=\linewidth]{"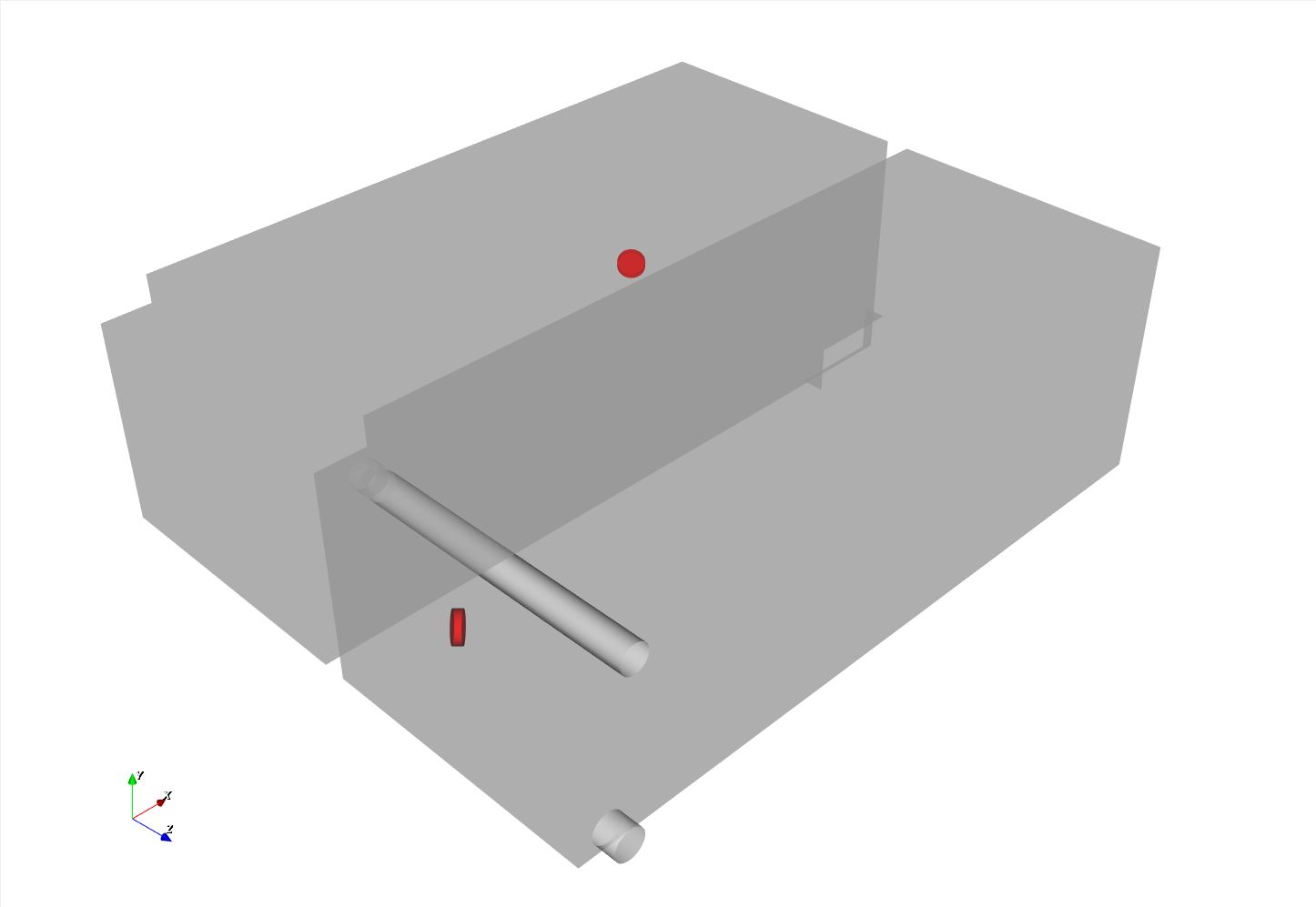"}
    \caption{Submersible mixers before meshing.}
    \label{fig:case2-equip}
    \end{subfigure}%
    ~ 
    \begin{subfigure}[t]{0.5\textwidth}
    \centering
    \includegraphics[trim={9.5cm 1.5cm 5cm 1.5cm}, clip, width=\linewidth]{"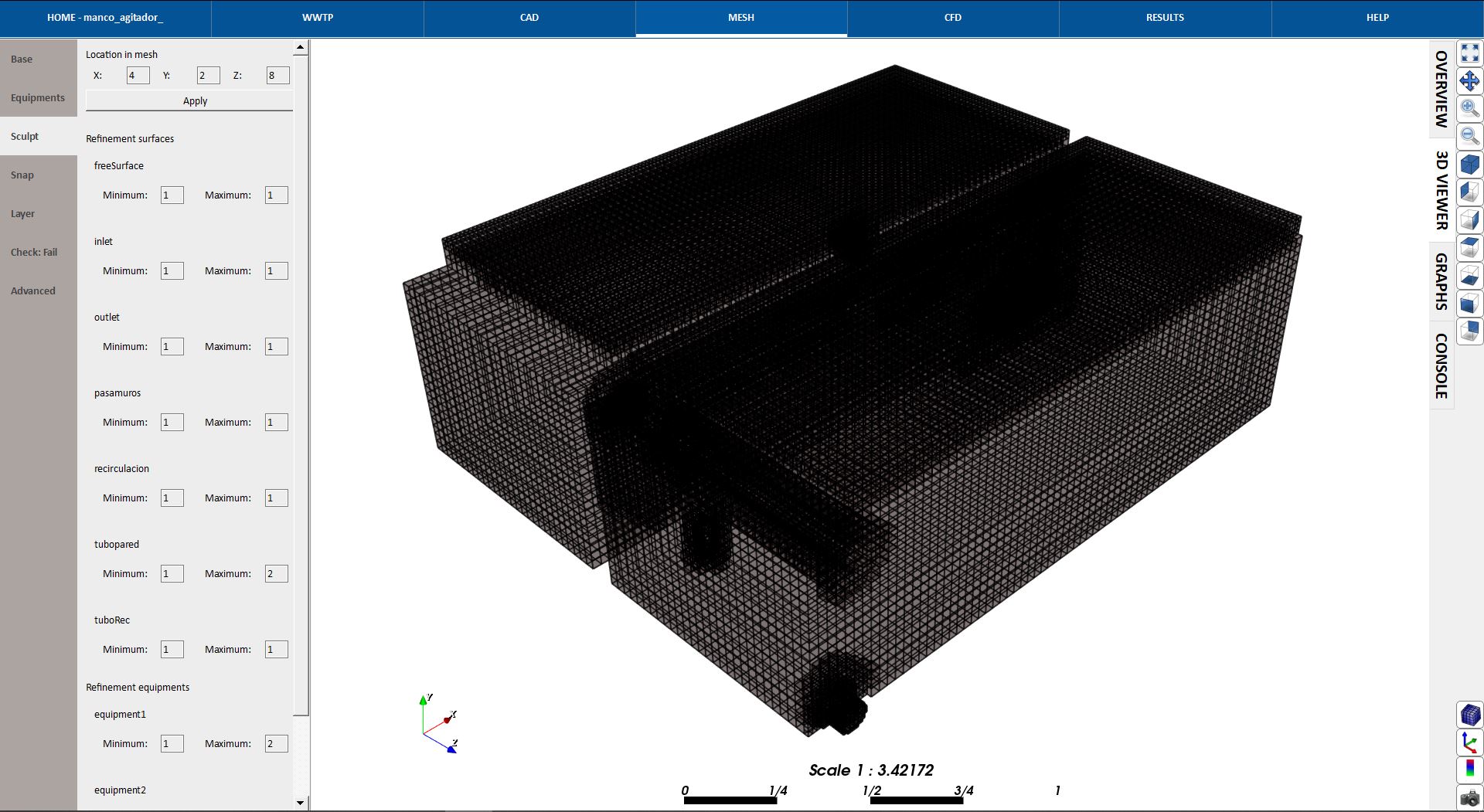"}
    \caption{Mesh with submersible mixers.}
    \label{fig:case2-mesh}
    \end{subfigure}
    \caption{Overviews of the Anoxic reactor under study.}
    \label{fig:case2}
\end{figure*}

HydroSludge provides a tool for easily including equipment, such as mixers into the CFD model (see Section~\ref{subsubsec:submersible}), so that two submersible mixers are inserted into the reactor. The new equipment can be configured with the following options: identification; center point; vertical inclination; azimuth inclination; radius size, and; full width. Figure~\ref{fig:case2-equip} showcases both created mixers inserted into the structure before meshing.

Once the surfaces, boundaries, and equipment are defined, it is time to proceed to the meshing procedure and the refinement of the resultant mesh, if needed.
Figure~\ref{fig:case2-mesh} contains the representation of the mesh with refinement at the submersible mixers.

OpenFOAM solver \textit{pimpleFoam} is selected for the simulation of the hydrodynamics taking place inside the anoxic reactor. The configuration mechanism is similar to the previous study case but, each solver has its intrinsic parameters which HydroSludge is responsible to present to the user. Concretely, this solver only needs the configuration of the rheological model and the flow rate (or velocity) for each inlet or outlet boundary. \textit{PimpleFoam} sets water as the default Newtonian model. % but Non-newtonian models (Ostwald de Waele and Hershel-Bulkley) are also available.
\par
Since this case counts with submersible mixers, their simulation features have to be defined. Specifically, HydroSludge expects the \textit{Manufacturer Coefficient}, and \textit{Power} for each impeller. \reviewMinor{Usually manufacturer coefficients are provided in clean water tests, hence it may be modified to match CFD results with real data measured in process water, as it is done in~\cite{Climent2018,Climent2019}.}

Finally, apart from the configuration of the simulation control, the execution can be parallelized (see Section~\ref{subsubsec:evaluation}). At the end of this step, the model will count with the hydrodynamic solution, which can be leveraged to apply transport equations for models such as the biological one described in the next section.

\subsubsection{Biological Model}\label{subsubsec:biological}
To apply the biological model implemented in the solver \textit{ASM1Foam} (see Section~\ref{subsubsec:solvers}), a time-step of the previous hydrodynamic solution has to be chosen. In this regard, the transport equation of the biological model will be applied to the selected time and all the data related to the mesh or submersible mixers are automatically shifted to this stage.
If HydroSludge does not detect hydrodynamic results, \textit{ASM1Foam} cannot be calculated. 

Next, the kinetic stoichiometric parameters of the ASM1 model must be defined. By default, HydroSludge loads the values established in ~\citep{henze1987}.
Additionally, the concentrations of all ASM1 state variables are set for every inflow boundary, as well as, the initial conditions of the reactor. After initialization, the case is ready to be simulated.

\subsubsection{Post-processing}
\begin{figure*}
    \centering
    \begin{subfigure}[t]{0.5\textwidth}
    \centering
    \includegraphics[trim={10cm 2cm 3cm 6.75cm}, clip,
    width=\linewidth]{"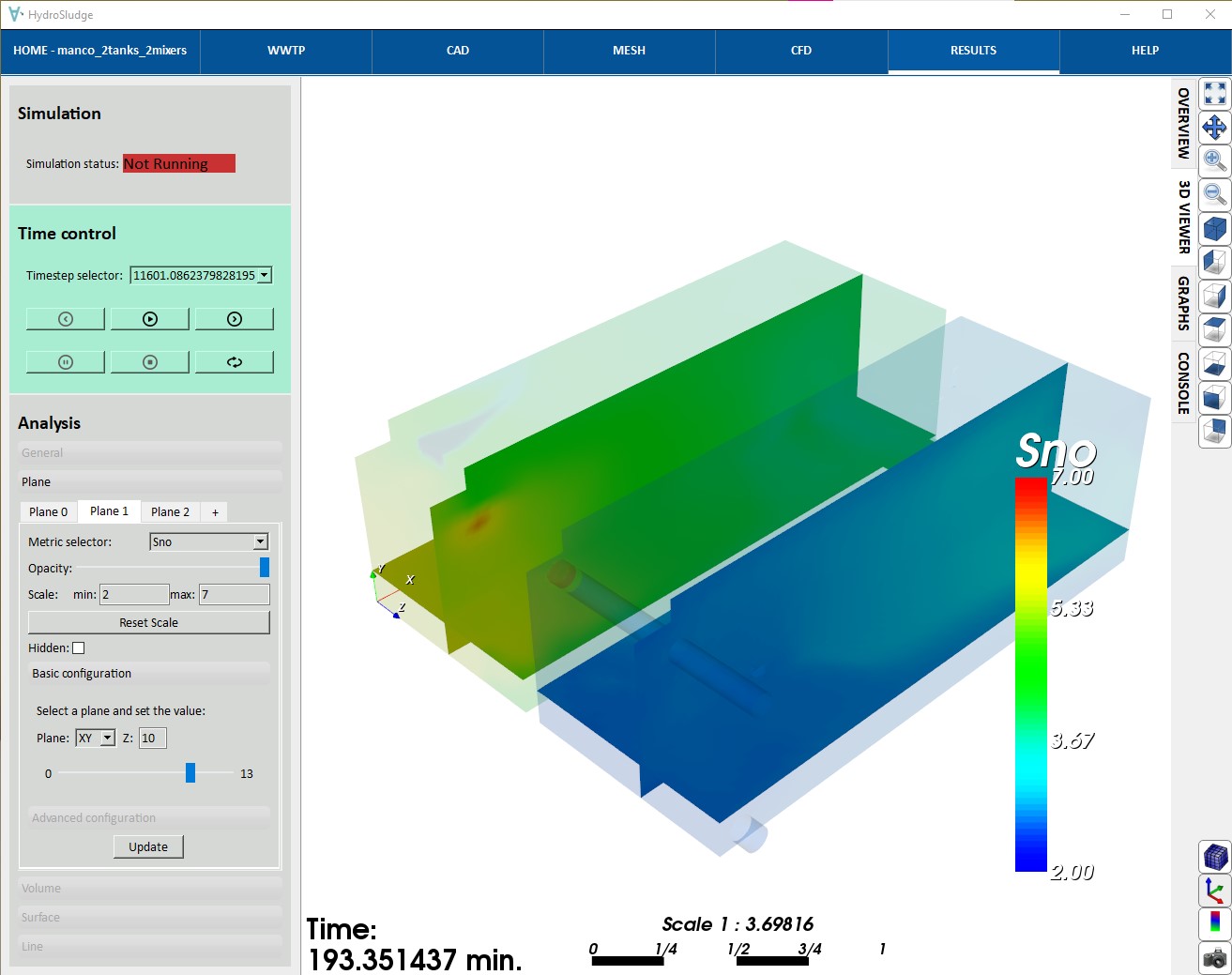"}
    \caption{Nitrate distribution in a plane.}
    \label{fig:case2-res1}
    \end{subfigure}%
    ~ 
    \begin{subfigure}[t]{0.5\textwidth}
        \centering
    \includegraphics[trim={8.6cm 0.5cm 0cm 5.75cm}, clip, width=\linewidth]{"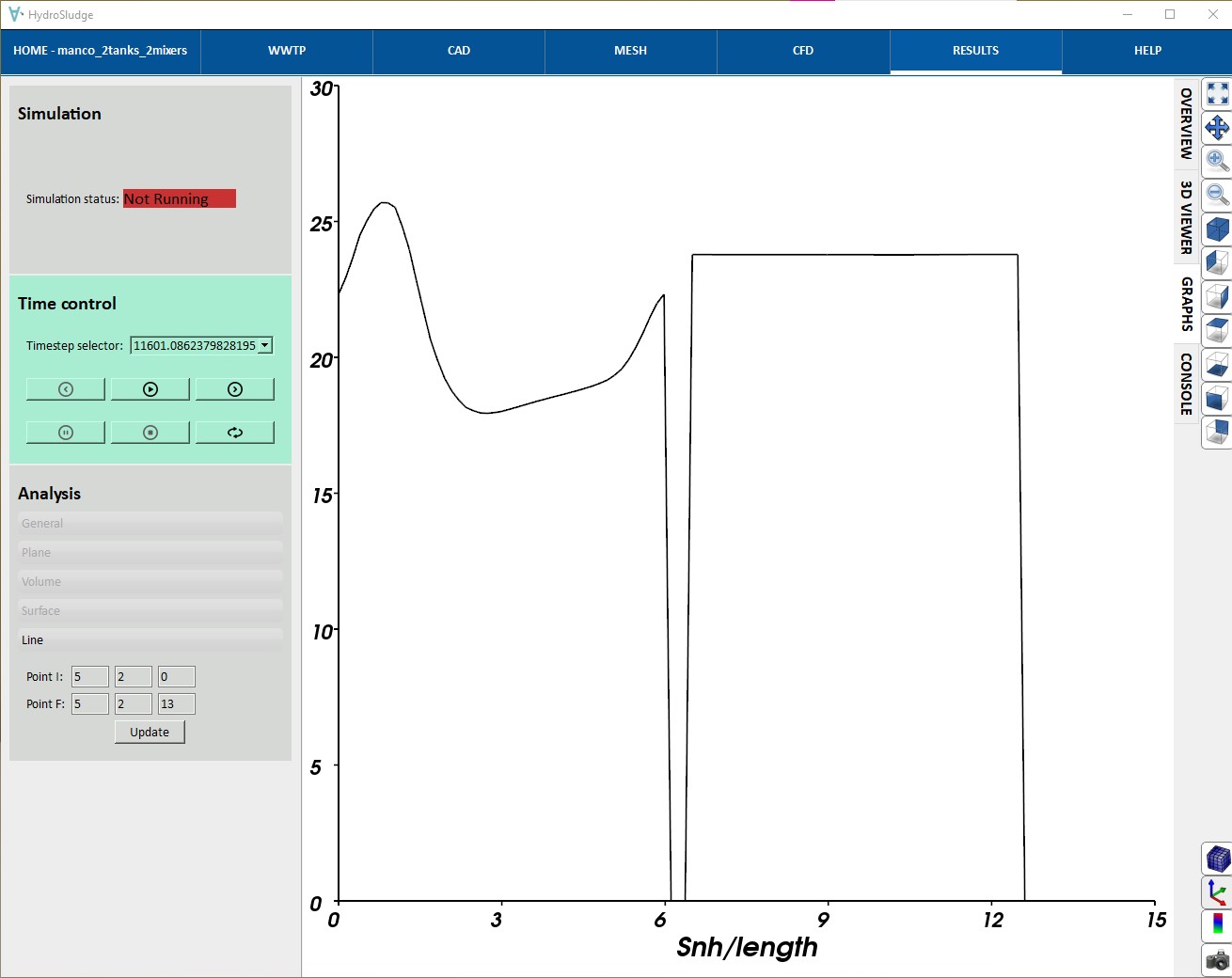"}
        \caption{Ammonia distribution along a probe.}
        \label{fig:case2-res2}
    \end{subfigure}
    
    \caption{Post-processing result examples.}
    \label{fig:case2-res3}
\end{figure*}

As soon as the CFD simulation yields results, users can explore them interactively.  \reviewMinor{For instance, Figure~\ref{fig:case2-res1} shows three planes coloring ``Sno'' variable for the $timestep\simeq11601.1$ that corresponds to the second $\simeq193.4$ of the transient simulation.} Another example of the tools that HydroSludge provides is a probe-line which is deployed inside the domain to study the behavior of a given variable. Figure~\ref{fig:case2-res3} depicts ``Snh'' distribution along a line in the Z dimension. Furthermore, the viewer and graph can also show the evolution in time of the variable thank to the media commands for playing the sequence.

\begin{figure}
    \centering
    \begin{subfigure}[t]{0.32\textwidth}
    \centering
    \includegraphics[trim={0.5cm 1.44cm 4.365cm 1cm}, clip,
    width=\linewidth]{"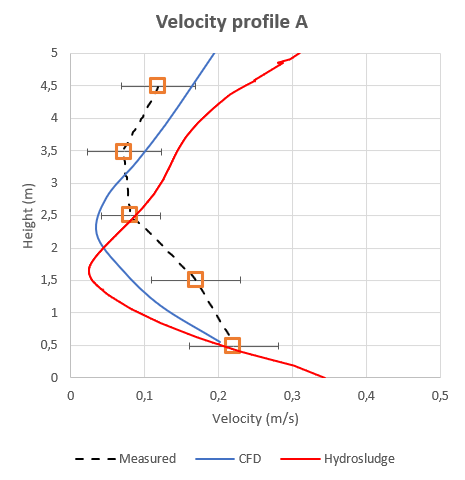"}
    \caption{Location A.}
    \label{fig:VelocityProfile_A}
    \end{subfigure}
    \begin{subfigure}[t]{0.229\textwidth}
        \centering
    \includegraphics[trim={1.74cm 1.42cm 5.2cm 1cm}, clip, width=\linewidth]{"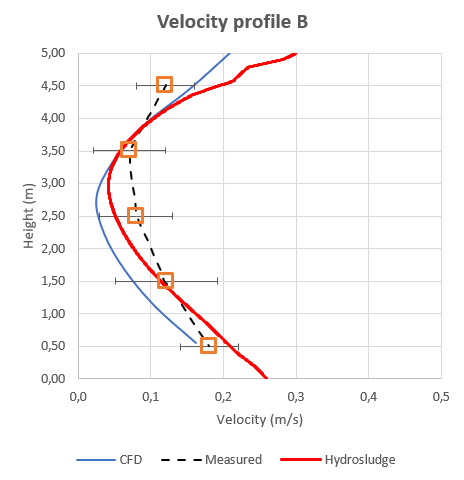"}
        \caption{Location B.}
        \label{fig:VelocityProfile_B}
    \end{subfigure}
    \begin{subfigure}[t]{0.272\textwidth}
        \centering
    \includegraphics[trim={1.75cm 1.38cm 4.3cm 1cm}, clip, width=\linewidth]{"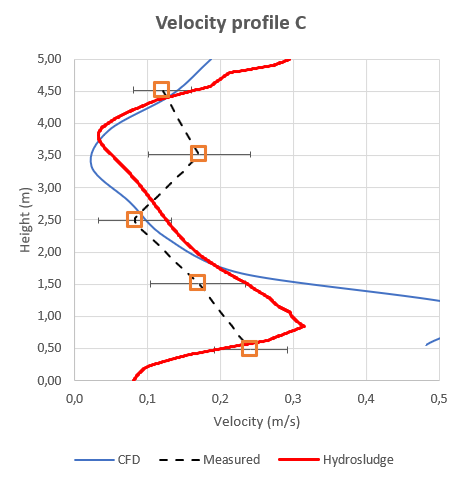"}
        \caption{Location C.}
        \label{fig:VelocityProfile_C}
    \end{subfigure}
    
    \caption{Velocity profile results at three different locations A, B, and C inside the Anoxic reactor; in black the measured data at the WRRF, in blue the results obtained by CFD, and in red the results obtained by Hydrosludge.}
    \label{fig:case2-res4}
\end{figure}

\review{Lastly, in order to validate experimental data collected at the WRRF, Figure~\ref{fig:case2-res4}  depicts the velocity profiles at three different locations obtained with Hydrosludge and ANSYS 19.2. The results obtained by Hydrosludge showed a similar adjustment to the data measured at the WRRF.}

\section{Conclusions}\label{sec:conclusions}
This paper, describes, implements, and evaluates a set of tools for the specific case of the study and analysis of wastewater treatment processes. \review{The toolbox methods are included in a software platform, named HydroSludge, to facilitate the evaluation and simulation of secondary clarifiers and conventional activated sludge processes in a WRRF}.
Future work places HydroSludge in the cloud as a web application instead of as a desktop application as new research developments indicate~\cite{Costa-Majo2021}. \reviewMinor{HydroSludge is also opened to be extended with other type of mixers such as aerators, vertical shaft or floating surfaces. Likewise, other ASM-like models can be incorporated in further versions.}

\reviewMinor{HydroSludge is a complementary WRRF modeling tool to Biowin, GPS-X, etc. Users may continue using plant-wide modeling to reproduce the entire operation of the facility.}

\review{HydroSludge is based on one of the most widespread and leading CFD simulation software, OpenFOAM.} However, HydroSludge prevents the user from getting involved in a complex process of configuring an extensive list of files, bringing a more wizard-like and intuitive usage. This feature makes HydroSludge an ideal tool for wastewater treatment industry users with low knowledge of CFD. Furthermore, HydroSludge includes the nomenclatures and procedures names used in the water sector.

\section*{Acknowledgements}
This work was supported by project RTC-2016-4560-7 from MINECO.
Researcher S.~Iserte was supported by the postdoctoral fellowship APOSTD/2020/026 from GVA and ESF. 
%Tirant III cluster of the \textit{Servei d'Informàtica} of the University of Valencia was leveraged in this research.
\review{Authors want to thank the anonymous reviewers whose suggestions significantly improved the quality of this manuscript.}
%\appendix
%\input{sections/appendixA}
\reviewMinor{
\section*{Software Availability}
\noindent\textbf{Software Name:} HydroSludge\newline
\noindent\textbf{Developers:} S. Iserte, D. Miralles, and C. Peña\newline
\noindent\textbf{System Requirements:} Windows 10\newline
\noindent\textbf{Program Language:} C++\newline
\noindent\textbf{Software Required:} Docker\newline
\noindent\textbf{Installer Availability:} On specific demand to siserte@uji.es\newline 
%\noindent\textbf{Installer Availability:} \url{https://bit.ly/2MbGwtZ}\newline 
%\noindent\textbf{Password:} HydroSludge2021\newline 
}

%\bibliographystyle{cas-model2-names}
\bibliographystyle{apacite}

\bibliography{cas-refs}

@inproceedings{Costa-Majo2021,
author = {Costa-Maj{\'{o}}, Francesc and Barreda, Paloma and Iserte, Sergio},
booktitle = {11th International Conference on Simulation and Modeling Methodologies, Technologies and Applications},
pages = {414--421},
title = {{A Distributed Mesh Generation Study Case through a Customizable Platform as a Service Framework}},
year = {2021}
}

@misc{Committee2014,
number = {518},
title = {{Recommended standards for Wastewater Facilities}},
url = {https://eec.ky.gov/Environmental-Protection/Water/PermitCert/Documents/10 State Standards - WW - 2014.pdf},
year = {2014}
}

@book{wef,
address = {US},
isbn = {0071464166},
publisher = {McGraw-Hill Professional},
title = {{Clarifier Design: WEF Manual of Practice No. FD-8}},
url = {https://mhebooklibrary.com/doi/book/10.1036/0071464166},
year = {2005}
}

@misc{cedex,
title = {{XXXVII Curso sobre Tratamiento de Aguas Residuales y Explotaci{\'{o}}n de Estaciones Depuradoras}},
url = {https://cpage.mpr.gob.es/producto/xxxvii-curso-sobre-tratamiento-de-aguas-residuales-y-explotacion-de-estaciones-depuradoras-2/},
year = {2019}
}

@article{Brennan2001,
   author = {Brennan, D.},
   issue = {January},
   journal = {Thesis},
   title = {The Numerical Simulation of Two-Phase Flows in Settling Tanks},
   year = {2001},
}

@article{Climent2018,
author = {Climent, J. and Basiero, L. and Mart{\'{i}}nez-Cuenca, R. and Berlanga, J. G. and Juli{\'{a}}n-L{\'{o}}pez, B. and Chiva, S.},
issn = {13858947},
journal = {Chemical Engineering Journal},
title = {{Biological reactor retrofitting using CFD-ASM modelling}},
volume = {348},
year = {2018}
}

@article{Climent2019,
author = {Climent, J. and Mart{\'{i}}nez-Cuenca, R. and Carratal{\`{a}}, P. and Gonz{\'{a}}lez-Ortega, M.J. and Abell{\'{a}}n, M. and Monr{\'{o}}s, G. and Chiva, S.},
issn = {1385-8947},
journal = {Chemical Engineering Journal},
publisher = {Elsevier},
title = {{A comprehensive hydrodynamic analysis of a full-scale oxidation ditch using Population Balance Modelling in CFD simulation}},
volume = {374},
year = {2019}
}

@article{Dapelo2018,
author = {Dapelo, D. and Bridgeman, J.},
journal = {Bioresource Tech.},
title = {{Assessment of mixing quality in full-scale, biogas-mixed anaerobic digestion using CFD}},
year = {2018}
}

@article{Ekama1997,
   author = {Ekama, G. A. and Barnard, J. L. and Günthert, F. W. and Krebs, P. and McCorquodale, J. A. and Parker, D. S. and Wahlberg, D. S.},
   journal = {Scientific and Technical Report},
   title = {Secondary Settling Tanks - Theory, Modelling, Design and Operation},
   year = {1997},
}

@article{Glover2006,
author = {Glover, G. C. and Printemps, C. and Essemiani, K. and Meinhold, J.},
issn = {02731223},
journal = {Water Science and Technology},
number = {3},
pages = {79--89},
pmid = {16605020},
publisher = {IWA Publishing},
title = {{Modelling of wastewater treatment plants - How far shall we go with sophisticated modelling tools?}},
volume = {53},
year = {2006}
}

@book{Grady2011,
address = {London},
author = {Grady, C. P. L. and Daigger, G. T. and Love, N. G. and Filipe, C.D.M.},
edition = {3rd Edition},
publisher = {IWA Publishing},
title = {{Biological Wastewater Treatment}},
year = {2011}
}

@article{henze1987,
author = {Henze, M. and Grady Jr, L. and Gujer, W. and Marais, G. and Matsuo, T.},
year = {1987},
title = {Activated Sludge Model No 1},
volume = {29},
journal = {Wat Sci Technol}
}

@article{Kynch1952,
   author = {Kynch, G. J.},
   issn = {00147672},
   journal = {Trans. Faraday Society},
   title = {A theory of sedimentation},
   year = {1952},
}

@book{JacekMakinia2010,
address = {London},
author = {{Makinia, J.}},
publisher = {IWA Publishing},
title = {{Mathematical Modelling and Computer Simulation of Activated Sludge Systems}},
year = {2010}
}

@article{Ovezea2009,
author = {Ovezea, A.},
journal = {Filtration and Separation},
number = {1},
pages = {24--27},
publisher = {Elsevier Advanced Technology},
title = {{Saving energy: Using fine bubble diffusers}},
volume = {46},
year = {2009}
}

@misc{Samstag2016,
author = {Samstag, R. W. and Ducoste, J. J. and Griborio, A. and Nopens, I. and Batstone, D. J. and Wicks, J. D. and Saunders, S. and Wicklein, E. A. and Kenny, G. and Laurent, J.},
booktitle = {Water Science and Technology},
number = {3},
pages = {549--563},
pmid = {27508360},
publisher = {Water Science Technology},
title = {{CFD for wastewater treatment: An overview}},
volume = {74},
year = {2016}
}

@book{Tchobanoglous2003,
author = {Tchobanoglous, G. and Burton, F. L. and Stensel, H. D. and {Metcalf \& Eddy.}},
isbn = {0070418780},
pages = {1819},
publisher = {McGraw-Hill},
title = {Wastewater engineering: treatment and reuse},
year = {2003}
}

@article{Takacs,
title = {A dynamic model of the clarification-thickening process},
journal = {Water Research},
volume = {25},
number = {10},
pages = {1263-1271},
year = {1991},
issn = {0043-1354},
author = {Takács, I. and Patry, G.G. and Nolasco, D.},
}

@article{vesilind1968design,
author = {Vesilind, P. A.},
journal = {Water Sewage Works},
number = {7},
pages = {302--307},
title = {Design of prototype thickeners from batch settling tests},
volume = {115},
year = {1968}
}
%\nocite{*}


%\include{sections/bio}

%\url{https://www.esi-group.com/products/cfd-multiphysics}
%\url{https://openfoam.com/products/visualcfd.php}
%simscale

\end{document}